\documentclass[a4paper]{jpconf}
\usepackage{graphicx}

\begin{document}

\title{Compass model on a ladder and square clusters}

\author{  Wojciech Brzezicki$^1$ and Andrzej M. Ole\'s$^{1,2}$  }

\address{$^1$ Marian Smoluchowski Institute of Physics, Jagellonian University,
              Reymonta 4,\\ PL-30059 Krak\'ow, Poland }
\address{$^2$ Max-Planck-Institut f\"ur Festk\"orperforschung,
              Heisenbergstr. 1, D-70569 Stuttgart, Germany }

\ead{w.brzezicki@gmail.com; a.m.oles@fkf.mpg.de}

\begin{abstract}
We obtained exact heat capacities of the quantum compass model on
the square $L\times L$ clusters with $L=2,3,4,5$ using Kernel
Polynomial Method and compare them with heat capacity of a large
compass ladder. Intersite correlations found in the ground state
for these systems demonstrate that the quantum compass model
differs from its classical version.
\end{abstract}


Quantum compass model originating from the orbital
(Kugel-Khomskii) superexchange in transition metal oxides has been
recently studied both using analytical \cite{Dou05} and numerical
\cite{Mil05,Wen08} approach. In spite of remarkable progress in
numerical methods for two--dimensional (2D) Ising--like models
\cite{Vid07,Cin08}, exact solutions are necessary to investigate
the structure of excited states. The one--dimensional (1D) compass
model is exactly solvable by mapping to quantum Ising model
\cite{Brz07} and exhibits interesting properties while approaching
to the quantum critical point at zero temperature \cite{Eri09}. In
addition, its ladder version, first considered by Dou\c{c}ot et
al. \cite{Dou05} is solvable in a similar way and its partition
function \cite{Brz09} can be obtained exactly in case of a large
(but finite) system. There is no exact solution for the 2D compass
model but the latest Monte Carlo data \cite{Wen08} prove that the
model exhibits a phase transition at finite temperature both in
quantum and classical version with symmetry breaking between $x$
and $z$ part of the Hamiltonian. This paper suggests a scenario
for a phase transition with increasing cluster size by the
behavior of the specific heat obtained via Kernel Polynomial
Method (KPM) \cite{Feh08}.

The Hamiltonian of the quantum compass model on a square $L\times
L$ lattice is given by
\begin{equation}
{\cal H}( \alpha )=(-1)^L J \sum_{i,w=1}^{L} \left\{ (1-\alpha)
\sigma ^x_{i,w} \sigma ^x_{i+1,w} + \alpha \sigma ^z_{i,w} \sigma
^z_{i,w+1} \right\} \label{ham1} ,
\end{equation}
where $\sigma^{z,x}_{i,w}$ are the $z$ and $x$ Pauli matrices for
site $\{i,w\}$, where $i$ ($w$) is a vertical (horizontal) index,
and we implement periodic boundary conditions. The sign factor
$(-1)^L$ is introduced to provide comparable ground state
properties for odd and even systems. Parameter $\alpha\in [0,1]$
makes this model interpolate between the situation when we have
$L$ independent Ising chains interacting with $x$ and $z$
components of spins for $\alpha=0$ and $\alpha=1$ respectively.
The case we have already discussed is the compass model on a
ladder \cite{Brz09}; this can be included into present discussion
by restricting the range of the summation over $i$ to the value
$2$ in the Hamiltonian.



The latter case is especially interesting, as we can obtain exact
ground state for any size $L$ of the system. The quantities
referring to the full spectrum, like the density of states, heat
capacity and thermodynamic correlation functions can be determined
for $L$ sufficiently big (like $L=52$) to be representative for
the thermodynamic limit. Solution for a ladder system is based on
the construction of invariant subspaces which are related to the
symmetry operators $\sigma^z_{1,w} \sigma^z_{2,w}$. It brings us
to a purely 1D Hamiltonian describing an Ising chain in transverse
field but depending on in which subspace we are --- some of the
$\sigma^z_{i}\sigma^z_{i+1}$ bonds are missing. An analogous
construction is possible for a square lattice but in this case
simplifications are rather modest; we cannot find an exact
solution anyway. This is not the case for a finite system - exact
diagonalization (ED) methods can be applied and symmetries can
reduce the Hilbert space considerably.

Ground state energies and energy gap of the model given by
(\ref{ham1}) has been already calculated for different values of
$\alpha$ and for $L\in[2,5]$ using ED and for higher $L$ using
Green's function Monte Carlo method \cite{Mil05}). Our approach
will be based on KPM \cite{Feh08} which will let us calculate the
densities of states and the partition functions for square
lattices of the sizes up to $L=5$. We start by applying Lanczos
algorithm to determine spectrum width which is needed for KPM
calculations. The resulting few lowest energies that we get from
the Lanczos recursion can be compared with the density of states
to check if the KPM results are correct. One should be aware that
the spectra of odd systems are qualitatively different from those
of even ones. For the even systems operator $S$ defined as
%
$S=\prod_{i,w=1}^{L} \frac12 \{1-(-1)^{i+w}\}\sigma^y_{i,w}$,
%
anticommutes with the Hamiltonian (\ref{ham1}). This means that
for every eigenvector $|v \rangle $ satisfying ${\cal H}
(\alpha)|v \rangle = E(\alpha) |v \rangle $ we have another
eigenvector $|w \rangle =S|v \rangle $ that satisfies ${\cal
H}(\alpha)|w \rangle = -E(\alpha) |w \rangle $. This proves that
even values of $L$ spectrum of ${\cal H}(\alpha)$ is symmetric
around zero but for odd $L$'s this does not hold; $S$ no longer
anticommutes with the Hamiltonian. To obtain symmetric spectrum we
would have to impose open boundary conditions.

We would like to highlight that KPM calculation for $5\times 5$
lattice ($2^25$ - dimensional Hilbert space) would be impossible
without using the symmetry operators. These operators are usually
called $P_i$ and $Q_w$ \cite{Dou05} and are defined for
$i,w=1,2,\cdots,L$ as
\begin{equation}
P_i=\prod_{w=1}^L\;\sigma^x_{i,w},  \hskip 1 cm Q_w=\prod_{i=1}^L
\;\sigma^z_{i,w}.
\end{equation}
One can easily check that although both operators commute with the
Hamiltonian, $P_i$ and $Q_w$ anticommute. Thus, we cannot find a
common eigenbasis for  $P$'s,  $Q$'s and the Hamiltonian; we
should find a different set of operators. A good choice is to take
all $P_i$ with $i=1,2,\dots,L$ and $R_w\equiv Q_w Q_{w+1}$ with
$w=1,2,\dots,L-1$. This gives us $(L-1)\times L$ commuting
symmetries. Let's denote their eigenvalues as $p_i$ and $r_w$
taking pseudospin values $\pm 1$. Each choice of $p_1,p_2,\dots
,p_L$ and $r_1,r_2,\cdots,r_{L-1}$ defines an invariant subspace
in which the Hamiltonian can be written in terms of $(L-1)\times
(L-1)$ new spin operators and $(L-1)\times L$ parameters $\{p_i\}$
and $\{r_w\}$. This statement can be proved by giving the explicit
form of the spin transformation.

The main benefit for us is that after the transformation the
Hamiltonian of $5 \times 5$ compass model ($\alpha=1/2$) turns
into $2^9$ spin models, each one on $4 \times 4$ lattice. In fact,
the number of {\it different} models is much lower than $2^9$;
most of resulting Hamiltonians differ only by a similarity
transformation. We can check this using Lanczos algorithm; if the
two lowest energy levels from two subspaces are the same then it
is reasonable to assume that these spectra are identical and the
Hamiltonians are the same. Finally, we find out that only $10$ out
of $512$ Hamiltonians are different; their two lowest energies and
degeneracies ars given in table \ref{ene10}. In fact, these
energies are known with much higher precision ($10^{-6}$) than
that given in the table 1, and we also get quite good estimation
for the highest energies. This gives us a starting point for KPM
calculations.
\begin{figure}[t!]
\includegraphics[width=14 cm]{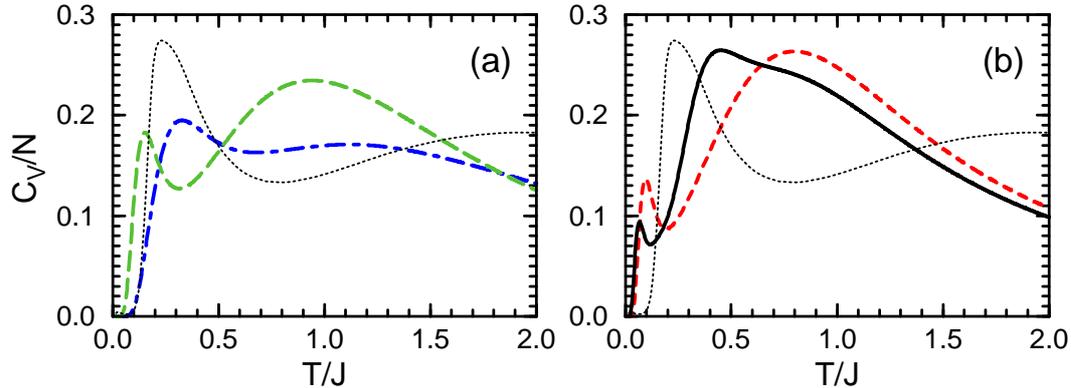}
\caption{Heat capacities as functions of temperature for
$\alpha=\frac{1}{2}$ for the compass clusters $L\times L$ of sizes
$L=2,3$ (panel (a), dashed-dotted and long-dashed lines) and
$L=4,5$ (panel (b), dashed and solid lines) and for the compass
ladder of $104$ spins (black dotted line).} \label{heat}
\end{figure}

\begin{table}[b!]
\caption{\label{ene10} Ground state energy $E_0$ and first excited
state energy $E_1$ and their degeneracies $d$ for $10$
nonequivalent subspaces of the $5\times 5$ compass model (1) at
$\alpha=1/2$. }
\begin{center}
\begin{tabular}{ccccccccccc}
\br
$n\!$ &1&2&3&4&5&6&7&8&9&10 \\
\mr
$\!E_0\!\!$
&$-14.54\!$&$-14.31\!$&$-14.30\!$&$-14.22\!$&$-13.75\!$&
$-13.67\!$&$-13.52\!$&$-13.45\!$&$-13.22\!$&$-12.79\!\!$ \\
$\!E_1\!\!$
&$-13.80\!$&$-13.15\!$&$-12.91\!$&$-12.50\!$&$-12.86\!$&
$-12.99\!$&$-13.26\!$&$-12.67\!$&$-12.88\!$&$-12.30\!\!$ \\
\mr
$d\!$ &2&20&20&20&50&100&50&100&100&50 \\
\br
\end{tabular}
\end{center}
\end{table}
Kernel Polynomial Method is based on the expansion into the series
of Chebyshev polynomials \cite{Feh08}. Chebyshev polynomial of the
$n$-th degree is defined as $T_n (x) = \cos [n \arccos x] $ where
$x \in [-1,1]$ and $n$ is integer. Further on we are going to
calculate $T_n$ of the Hamiltonian so first we need to renormalize
it so that its spectrum fits the interval $[-1,1]$. This can be
done easily if we know the width of the spectrum. Our aim is to
calculate the renormalized density of states $\tilde{\rho} (E)$
given by
%
$\tilde{\rho}(E)=(1/D)\sum_{n=0}^{D-1}\delta (E-\tilde{E_n})$,
%
where the sum is over eigenstates of ${\cal H}(\alpha)$ and $D$ is
the dimension of the Hilbert space. The moments $\mu_n$ of the
expansion of $\tilde{\rho} (E)$ in basis of Chebyshev polynomials
can be expressed by
\begin{equation}
\mu_n= \int_{-1}^{1} T_n(E) \tilde{\rho} (E) dE =\frac{1}{D}
\Tr\{T_n(\tilde{\cal{H}})\}\,.
\end{equation}
Trace can be efficiently estimated using stochastic approximation:
\begin{equation}
\Tr\,\{T_n(\tilde{\cal{H}})\} \approx \frac{1}{R}\sum_{r=1}^R
\;\langle r|T_n(\tilde{{ \cal H}})|r \rangle\,,
\end{equation}
where $|r \rangle$ ($r=1,2,\dots,R$) are randomly picked complex
vectors with components $\chi_{r,k}$ ($k=1,2,\dots,D$) satisfying
$ \langle \chi_{r,k} \rangle = 0$,  $\langle \chi_{r,k}\chi_{r',l}
\rangle = 0$, $\langle \bar\chi_{r,k}\chi_{r',l} \rangle =
\delta_{r,r'}\delta_{k,l}$ (average is taken over the probability
distribution). This approximation converges very rapidly to the
true value of the trace, especially for large $D$. Action of the
$T_n(\tilde{\cal{H}})$ operator on a vector $|r\rangle $ can be
determined recursively using the following relation between
Chebyshev polynomials:
%
$T_n(\tilde{\cal H})| r \rangle = \{2\tilde{\cal H}\,
T_{n-1}(\tilde{\cal H})-T_{n-2}(\tilde{\cal H})\} |r\rangle$.
%
We can also use the relation $2T_{m} (x)T_{n} (x) = T_{m+n} (x) +
T_{m-n}(x)$ to get moments $\mu_{2n}$ from the polynomials of the
degree $n$. Finally the required function,
\begin{equation}
\tilde{\rho} (E)\approx \frac{1}{\pi \sqrt {1-E^2}} \left\{g_0 \mu
_0 + 2\sum_{n=1}^{N-1}g_n \mu _n T_n(E) \right\}\,,
\end{equation}
can be reconstructed from the $N$ known moments, where $g_n$
coefficients come from the integral kernel we use for better
convergence. Here we use Jackson kernel. Choosing the arguments of
$\tilde{\rho} (E)$ as being equal to $E_k=\cos [(2k-1)\pi/2N']$
($k=1,2,\dots,N'$) we can change the last formula into a cosine
Fourier series and use Fast Fourier Transform algorithms to obtain
rapidly $\tilde{\rho} (E_k)$. This point is crucial when $N$ and
$N'$ are large, which is the case here; our choice will be
$N=20000$ and $N'=2N$. In such a way we can get the density of
states for $4\times 4$ and $5\times 5$ systems. In the latter case
we obtain $10$ energy spectra for $10$ nonequivalent subspaces
--- these can be summed up with proper degeneracy factors (see
table \ref{ene10}) to get the final density of states
$\tilde{\rho}(E)$ and next the partition function via rescaling
and numerical integration.

\begin{figure}[t!]
\includegraphics[width=7.1 cm]{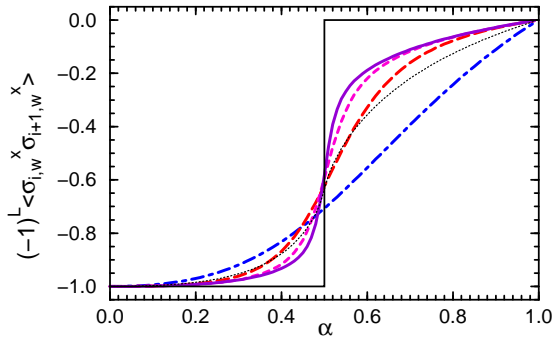}\hspace{3pc}
\begin{minipage}[b]{7.5 cm}
\caption{Correlations $\langle\sigma^x_{i,w}\sigma^x_{i+1,w}
\rangle$ in the ground state calculated for sizes $L=2,3,4,5$
(dashed-dotted, long-dashed, dashed and solid line) and for
infinite ladder (black dotted line) as functions of $\alpha$. Step
function shows a discontinuous transition for a classical compass
model at $\alpha=1/2$. } \label{cor}
\end{minipage}
\end{figure}


The heat capacity $C_V$ for the compass $L\times L$ clusters
behaves differently from that for a compass ladder, see
\fref{heat}. The main difference is vanishing of the low--$T$ peak
when the system's size increases. This correspond to vanishing of
the low--energy modes which is consistent with presence of the
ordered phase for finite $T$ in the thermodynamic limit. In
contrast, heat capacity of the ladder indicates robust low--energy
excitations and dense excitation spectrum at higher energies
causing a broad peak in $C_V$.

In \fref{cor} we compare nearest neighbor correlations as
functions of $\alpha$ for finite clusters, infinite compass ladder
and classical compass model on a square lattice. Curves for finite
clusters converge to certain final functions which is something
intermediate between classical case and quantum ladder. This
result shows that even in large $L$ limit the 2D compass model
preserves quantum correction to a classical behavior even though
it chooses ordering in one direction \cite{Wen08}.

In summary, we have shown that exact heat capacities of square
compass $L\times L$ clusters could be obtained by implementing the
symmetries up to $L=5$. The behavior of the low--$T$ peak in the
heat capacity indicates that the gap in the spectrum decreases
with increasing $L$. This agrees with the numerical results
obtained before by numerical approach \cite{Mil05}.

\ack
We acknowledge support by the Foundation for Polish Science
(FNP) and by the Polish Ministry of Science and Higher Education
under Project No.~N202 068 32/1481.





\end{document}